# Energy Estimates Across Layers of Computing:
## *From Devices to Large-Scale Applications in Machine Learning for Natural Language Processing, Scientific Computing, and Cryptocurrency Mining*[1]

Sadasivan Shankar
SLAC National Laboratory, Menlo Park, CA; Materials Science and Engineering, Stanford University, CA, USA
{sshankar@slac.stanford.edu; sadas.shankar@stanford.edu}

*Abstract*—Estimates of energy usage in layers of computing from devices to algorithms have been determined and analyzed. Building on the previous analysis [3], energy needed from single devices and systems including three large-scale computing applications such as Artificial Intelligence (AI)/Machine Learning for Natural Language Processing, Scientific Simulations, and Cryptocurrency Mining have been estimated. In contrast to the bit-level switching, in which transistors achieved energy efficiency due to geometrical scaling, higher energy is expended both at the at the instructions and simulations levels of an application. Additionally, the analysis based on AI/ML Accelerators indicate that changes in architectures using an older semiconductor technology node have comparable energy efficiency with a different architecture using a newer technology. Further comparisons of the energy in computing systems with the thermodynamic and biological limits, indicate that there is a 27-36 orders of magnitude higher energy requirements for total simulation of an application. These energy estimates underscore the need for serious considerations of energy efficiency in computing by including energy as a design parameter, enabling growing needs of compute-intensive applications in a digital world.

## I. Introduction

As computing becomes ubiquitous from intelligent sensing at the edge to digitalization of systems and with wider adoption of Artificial Intelligence and Machine Learning (AI/ML), it is clear that energy used in computing is expected to increase non-linearly [1, 2]. In continuation of the earlier analysis [3], in this work, additional quantitative analysis to include hardware components from the transistor level to including the Application have been provided, where 'Application' is defined to include the entire simulation by the process of computation.

Energy for Application is a metric that includes algorithms and software similar to energy/instructions at the hardware system level and energy/bit switching at the transistor level. Specific examples of computational simulations included in this analysis are training of a large language model (LLM) or inference needed for machine learning applications [5], large-scale simulation of a single Covid virion particle [6], and computer mining of a single Crypto coin.

The previous analysis illustrated that energy efficiency in computing decreases as one goes up in layers from the bit switching of a single device to instruction at the system level, and included an example of the entire simulation of the Application [3]. As will be illustrated with more examples below, the algorithms implemented in a software require significant energy for the computation. The increase in energy usage through the layers is due to the complex trade-offs in implementing the computing architecture in hardware, and in implementation of algorithms in software.

In Section II, the energy needed for these operations are estimated using both top-down and bottom-up analyses. In the top-down analysis, the total energy required by the system and the number of *switchings* or operations are used to estimate the unit energy required. In the bottom-up analysis, published literature estimates are used to illustrate energy requirements for specific system components (e.g., energy for on-chip and off-chip communications, memory, etc.). These estimates are approximate, but help in understanding the increase in energy needs across the layers, from the transistor-level to the higher levels of computing.

In Section III, energy required for three large-scale computing applications are estimated: AI/Machine Learning in large-scale Natural Language Processing for training and for inference (e.g., ChatGPT), scientific simulation of Covid virion, and Cryptocurrency mining for a single coin based on analysis of published literature. Section IV provides energy for a system based on thermodynamics and biochemical analysis of metabolism in cells. These provide lower limits of energy determined by physics, chemistry, and biology.

In the final section V on Conclusions, combining all of the above analysis, an integrated picture of energy utilization spanning large orders of magnitudes from bits to applications are provided. Comparing these energy estimates of a few computing applications with physical and biological limits (from Section IV) provides a unique and compelling picture for why energy efficiency should be part of the design of microelectronics and computing in the future.

## II. Energy Trends in hardware and architectures

The energy analysis in this section is sub-divided into two parts: System-level (e.g., microprocessor) and Component-level (e.g., transistor switching, communication, memory) analysis. Building on the previous work [3,4], in the first sub-section II.A, the energy usage across technologies and architectures over a period of about ten years is compared. In the second sub-section II.B, the analysis of sub-components of information processing is used to understand how the energy requirement per unit operation increases as one goes up through the layers of computing from switches to systems.

### A. Hardware and Architectures

The energy trends in different architectures for AI/ML systems across three aspects are illustrated in Figure 1: Technology nodes, Architectures (between different systems), and different precisions in Instructions or Operations.

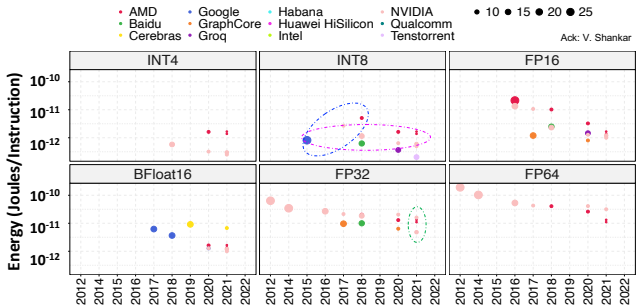

**Figure 1**: Energy estimated per operation for different processors (twelve in number) and different technologies (four in number) from 2012 to 2022 in Joules/Instruction, with different colors indicating different architectures from different companies. Smaller circles represent advanced technology nodes, while larger circles indicate older technologies. Three ellipses are drawn for comparisons: Blue: Older technology can be more energy efficient; Magenta: Small change in energy between technology and architecture; Green: Both technology node and architecture can provide comparable energy efficiencies. The x-axis is the year in which the benchmark was estimated, and the y-axis is the Energy in Joules/Instruction for different precisions: INT4 is 4-bit, INT8 is 8-bit, FP16, FP32, and FP64 are floating point instructions with 16-bit, 32-bit, and 64-bit respectively. BFloat16 is truncated (16-bit) version of the 32-bit IEEE 754 single-precision floating-point format (binary32) for AI/ML.

Besides architectures, there are other factors that are critical to the system energy, including integration within a chip or a package, the size and levels of memory, etc. One key observation is that technology scaling alone does not provide reduction in energy required for executing an instruction. To illustrate this, three ellipses are drawn in Figure 1, where the individual technologies and processors are represented by the size of the circles and their colors respectively (given at the top of the figure). Older technologies are represented as larger circles, while newer or smaller technologies as smaller circles.

The blue ellipse (for INT8, 8-bit integer operation) shows that a newer technology product may not be as energy efficient as a different architecture from an older technology. The pink ellipse indicates that neither the technology nor the architecture provides observable change in energy efficiency, while the green ellipse for 32-bit floating point instructions indicates that the processor on an older technology can be more energy efficient than an architecture on the newer technology.

### B. Components: Interconnects, Memory, Communication, and others

The previous analysis [3] indicated that energy efficiency of computing systems decreases as more components are integrated on top of the transistors for designing larger computing systems. There is a loss in energy efficiency as one goes up in the layers of computing (e.g., from transistors to memory cell to architecture). This can be partially attributed to data transfer including fetch and storage both at the chip level and beyond the chips in boards and systems. In order to analyze the features that drive this loss in energy efficiency, Figure 2 is based on published work [5-9] to compare energy per bit operation for information processing/communication. As shown in Fig. 2, the energy estimates range from the transistor level energy of $2.87 \times 10^{-19}$ joules for switching to $3.0 \times 10^{-5}$ joules for wireless transfer of a bit. This large disparity highlights the need for inclusion of energy needs to expand to computing systems including communication of data within the chip, across longer distances in systems, and beyond. Estimates on maximum DRAM access energy for given peak bandwidth and a given power budget of 60W indicate a range of $3.9 \times 10^{-12}$ joules to $1.4 \times 10^{-11}$ joules further indicating the consistency between the different sources in literature (order of picojoules) [10]. Taken together, these estimates reaffirm that storing, accessing data from high-speed memories, interconnects which are used to transport bits, off-chip communication, wireless transfer, and long-distance communication all appear to be more energy-intensive than a typical transistor switching. In addition, these provide pointers to the increase of energy requirements as one from goes up the layers of computing.

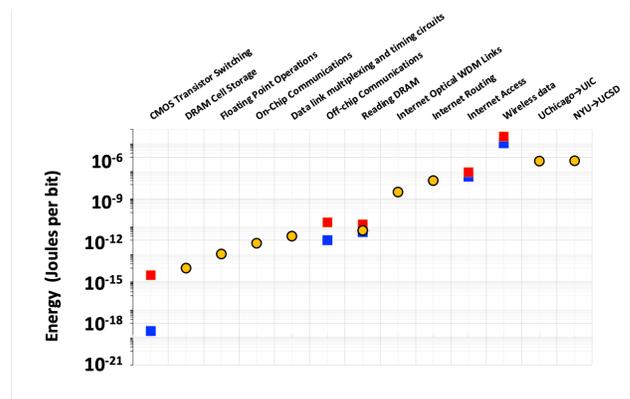

**Figure 2**: Energy as estimated per bit for different components of information processing taken from multiple sources [5-9]. In cases, where there are multiple estimates, we color them to indicate the bounds: Blue is the minimum estimate and red is the maximum estimate, while orange circles represent single point estimates. The energy estimates range from the transistor to long distance communication of data. Among the compared components, the wireless transfer appears to be the most energy-intensive followed by that for data transfer over long distances, with the bit-level switching at the transistor as the least energy-intensive.

## III. ENERGY ESTIMATES OF APPLICATIONS

Although hardware and architecture are significant components of integrated systems, algorithms and software are also critical for enabling computations as needed for applications. This link can be even be traced to the Church-Turing analysis of the halting problem. Mathematically, this analysis proved that no general algorithm exists that solves the halting problem for all possible program–input pairs independent of the hardware or architecture, thereby indicating the importance of specific algorithms and software used to simulate applications [11,12]. In addition, as will be illustrated below, the time to complete a simulation for the Application is most strongly influenced by the underlying algorithms implemented in the corresponding software.

In order to estimate the energy for a simulation, the following aspects are used in the analysis: First, the energy for the operation is based on floating point operations unless specified otherwise. These include Natural Language Processing (for training/inference) and scientific computations (FP32 or FP64). These are used to set baseline numbers ranging from $1 \times 10^{-11}$ to $1 \times 10^{-10}$ Joules per operation. Second, the total number of simulation cycles are estimated from the respective papers and publicly available literature and are explicitly summarized below. For inference and cryptocurrency mining, energy estimates used are from the respective publications.

In the first sub-section III.A below, energy for training and inferring Machine Learning models for Natural Language Processing (NLP) is estimated. In the second sub-section III.B, energy of a specific scientific simulation or computation is estimated, and in the last sub-section III.C, similar estimates are reported for a single crypto coin mining.

### A. AI/ML for Natural Language Processing

Within AI/ML applications, Nature Language Processing (NLP) is used to parse vast amounts of literature in all languages and also enable computer-aided translation between the different languages. These AI/ML methods depend on training on a large corpus, namely significant amounts of data using words, phrases, part-of speech requirements, existing collections of text from academic journals, books, social network websites, Wikipedia, and Common Crawl (crawling the web). Although not all of these are curated web sites, applications of current NLP methods for reproducing texts have achieved reasonable accuracies [13].

To estimate the training cost for these NLP algorithms, the energies based on the number of floating-point operations are used as in the precursor work [3]. Other analysis has assessed that the number of floating-point operations needed for large-scale models should scale with the number of tokens (a metric used for the training effort in these models) [14-16]. The initial training estimates for the number of floating-point operations range from $6 \times 10^{18}$ to $3 \times 10^{24}$, exceeding conservative estimates from the previous work [3]. As before, assuming the energy per instruction (EPI) to range from $1 \times 10^{-12}$ Joule/FPI (FP16) (lower bound) to $1 \times 10^{-11}$ (FP64) (upper bound) representing two cases, the energy required varies from 6 million joules to 30 trillion joules. Clearly, these highlight the large energy budgets required for training, consistent with published findings on the large energy costs for training these algorithms [17]. Although the energy cost of inference is spread over its usage, the time for inference is longer than the time for training. Retraining a large model is less compute-intensive, but given the power law nature of large language models [18] and new words and colloquial language regularly entering the corpus, the retraining can also become computationally energy-intensive over time.

For inference, the energy estimates are based on the usage patterns of Large Language Models [19-20]. The bounds are calculated assuming ten million queries/day, each query limited to five hundred words. The number of floating-point (FP) operations to compute an inference for a model with 175 billion parameters vary between 1 million FPs to 175 billion FPs per word. Appropriately, energy estimated ranges from 41 million joules (1 million FPs per word) to 798 billion joules (175 billion FPs) per year. The time duration for inference is longer (estimated here annually, given its regular estimated usage) compared to that for training (estimated to be about fifty-five days, once per year). At the higher bound, the energy used in inference is significantly higher than per capita electricity usage in the US and even higher than the total energy expended by a human in a lifetime of seventy-three years. These estimates are plotted in Figure 3.

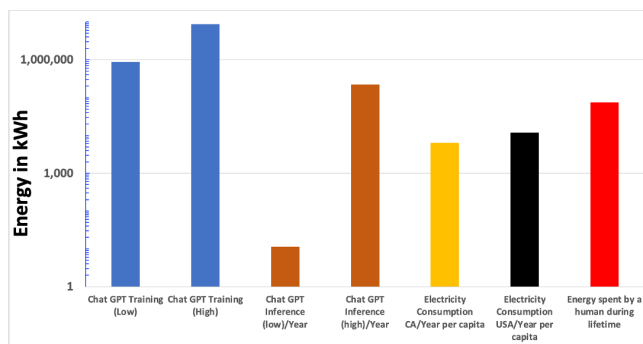

**Figure 3**: Energy in kilowatt-hour for training (during a single training activity of about fifty five days) and inference (over a period of the year) compared to per capita energy usage in the US and the total energy usage in a human's lifetime. The y-axis in logarithmic units show the significant energy needs for both training and inference.

### B. Scientific Computing using Supercomputers

In this section, we quantify the energy needed per simulation (defined as Application, as mentioned previously in Section I) based on a very large-scale scientific computation of a virion, specifically SARS-CoV-2 spike protein, the main viral infection machinery for Covid19 [21]. The simulation, which was run on a supercomputer, investigated the spike dynamics of SARS-CoV-2 viral envelope which consisted of 305 million atoms.

The energy estimates are based on the following parameters as indicated in the paper [21]. Specifically, simulations were run for 8.77 days on 80 P100 GPUs on Comet at San Diego Supercomputer Center, for sampling of ∼7.5 micro seconds. The simulation parameters are given in Table 1. Based on the simulation time and the total performance, and assuming that the energy per FP operation to be bounded by 1 x 10$^{-12}$ to 1 x 10$^{-11}$ joules, the energy for Spike ACE Complex and SARS Covid virion simulation is estimated to be about 24.9 billion joules (Table 2).

| NAMD Simulation | Atoms | Nodes | Sim rate | Performance |
|---|---|---|---|---|
| Spike-ACE2 complex | 8.5M | 1024 | 162 ns/day | 229 TFLOP/s |
| SARS-CoV-2 virion | 305M | 4096 | 68 ns/day | 3.06 PFLOP/s |

**Table 1**: Simulation parameters for Covid Virion particle simulations.

| Application | Energy (Joules) | Energy (kWhr) |
|---|---|---|
| Spike ACE Complex | 1.74E+09 | 4.82E+02 |
| SARS Covid Virion | 2.32E+10 | 6.44E+03 |
| TOTAL (Max) | 2.49E+10 | 6.92E+03 |

**Table 2**: Energy estimate in Joules and kWh for simulation of a single virion particle.

It can be observed that the energy needed for this entire simulation is higher, compared to the inference in the previous Application of a large language model, although the duration of the virion simulation was shorter compared to inference. The energy itself is over twenty orders of magnitude higher than the energy for a single floating-point instruction [3]. This because of the high power expended by a large scale-supercomputer and the significant number of compute cycles needed for scientific computations. Thus, a large-scale scientific computation based on detailed physics and chemistry also requires significant energy to complete the simulation compared with AI/ML analysis.

*C. Cryptocoin Mining*

In this section, energy estimate for cryptocoin mining is provided as another example of Application. The cryptocoin computing as needed for cryptocurrency mining is the result of computational operation of a virtual online transaction that under certain conditions is included in a ledger, which is validated by connected online computer nodes. The design of the operation is compute-intensive both due to the nature of the one-way operation termed as Hashing that maps digital inputs into a fixed length of output digits [22-23]. The digital operation of crypto coin mining has already been observed to be energy intensive that by estimates are close to 131 terawatt-hours annually exceeding the electricity of some countries and even requiring large dedicated installations for the mining computation including in the US [24].

Due to the variability in energy estimates to "mine" a single coin in a computer, two separate values are used to provide bounds: a lower limit of 661.36 kW-hours and an upper limit of 1449 kW-hours as energy needed to mine a single bitcoin. These energy estimates are plotted in Figure 5 for comparisons with other applications in the final section.

## IV. ENERGY LIMITS IN COMPUTING

All microscopic systems are subjected to thermodynamic fluctuations, which in turn limit energy of processes at a given temperature. It is especially critical to all systems in nature, specifically biological systems (as opposed to human-made) as they are subject to significant noise in the system, especially when they are connected to the environment. The noise in these systems is subjected to the laws of thermodynamics. On average, this noise is typically quantified in terms of the temperature of the system in Kelvin (K). These fluctuations are proportional to the average temperature (T) and is given by $k_BT$, where $k_B$ is the Boltzmann constant (1.38 x 10$^{-23}$ joules/K/molecule). For a temperature corresponding to 27 degrees Celsius (~80.6 degrees Fahrenheit, 300.15 Kelvin), the average thermodynamic energy is 4.14 x 10$^{-21}$ joules/molecule (*thermodynamic limit*).

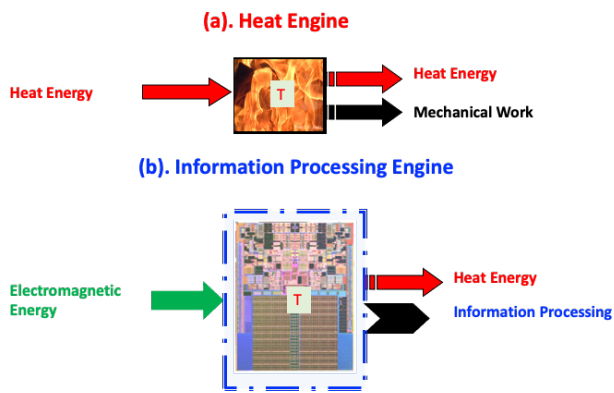

**Figure 4**: Energy partitioning comparisons between a thermodynamic heat engine and an information processing engine, represented as simple transduction systems. Both are subject to thermodynamical conditions, represented by the temperature T (in the figures). Computers fall under the second class.

All systems, both natural and synthetic, and their operations can be classified in terms of how the various forms of energy are interconverted between the forms including with heat, as part of a natural process or for producing useful work. Comparisons can be conducted between thermodynamic systems such as heat engines (which convert heat into mechanical work and the rest dissipated as lower quality heat) and information processing systems (which convert electromagnetic energy into information and then dissipated as heat). This is illustrated in Figure 4. Both operate at some finite temperature T, which may be uniform throughout or could vary depending on the local conditions. In addition, for information processing, there are various physical formalisms that build upon the above simple premise for determining the minimum energy of information processing. An example of this is the energy required for memory erasure is estimated to be $k_BT \log_e(2)$, where $\log_e(2)$ refers to the natural logarithm to the base 2 [25-26]. This has been further expanded to systems level for estimating the lower limits of gates and even macroscopic information processing systems [27-28]. In the current analysis, $k_BT$ is used to set lower bound of minimum energy of computing. This simple estimate provides a baseline for any microscopic entity involved in both information and physical processing (e.g., metabolism). The key point is that in natural systems, quantum-based processes, and in supercomputers, energy for any operation is subjected to this limit. Another important consequence of this limit is that it can go lower as in cryogenic systems or higher when systems operate at higher temperatures (e.g., steam engines, mammals).

Although we have indicated that the thermodynamic limit sets the lower bound of energy required for information processing, many of the natural, especially biological systems operate away from equilibrium needing higher energies. In order to estimate the limiting energy associated for such as a system (*biological limit*), we will use the fundamental metabolic reaction of ATP hydrolysis for storing energy as a benchmark. Adenosine triphosphate (ATP) is an organic compound, considered as the energy currency for driving metabolism in biological organisms. ATP is consumed for providing energy in processes including muscle contraction, nerve signaling, substrate phosphorylation, and biosynthesis of organelles [29]. The ATP stores higher energy in its phosphorous-oxygen [P-O] bonds and the extra energy is released in breaking the molecule down with fewer P-O bonds. Given the ubiquitous nature of this process as the basic unit of energy transfer, most of the known biological organisms use this mechanism for enabling information processing. For our estimates, we shall use the standard free energy of ATP hydrolysis (breaking of Phosphorous-Oxygen bonds) to be -31.55±1.27 kJ·mol at 298 K and pH of 7 [30]. This energy corresponds to $5.24 \times 10^{-20}$ Joules/molecule. Comparing this energy with the thermodynamic limit is illuminating as biological unit of energy is about an order of magnitude higher than the thermodynamic noise limit. All physical systems including computers need higher energy for information processing as they operate away from equilibrium conditions.

## V. CONCLUSIONS

As we have shown above, the energy used in computing continues to increase as we go from atomic level (molecule or nano-sized transistors) to systems level and goes even higher for a specific simulation of the Application. The layers of computing analyzed in this work include bit processing at the switch or transistor level, instructional operation at the system level, and Application at the simulation level. Further, the defined application is used as the basis of the metric to estimate the energy used by algorithms implemented in a relevant software. These estimates of the different layers of computing along with the physical and biological energy limits are all illustrated in Figure 5, which compares different energies in terms of joules per unit process. This process can be anything from thermodynamical fluctuation to a single transistor switching, or a single instructional operation, or executing a HPCG (High performance Conjugate Gradient) operation, to the entire simulation of the Application. The comparisons show over 27-36 orders of magnitude in energy increase per unit information processing in going from thermodynamical and biological limits to large-scale simulations. The most recent single transistors are less than three orders of magnitude higher than the thermodynamic noise limit.

Several other observations can be made from this analysis. (1) Given the large range of energy estimates, the algorithms and software provide the biggest areas for improvement on energy efficiency; (2) The energy for training of AI/ML systems seem to be significantly higher than inference. However, since inference is run over longer time, their cumulative energy usage can become comparable to that of training; (3) The system-level energy estimates are still over four orders of magnitude higher than that of single transistors, indicating improvement opportunities at the architecture and hardware level; (4) This analysis, consistent with other studies [5-10], indicates that data access and memory storage, data movement including through the local interconnects, and long-scale communications all require higher energy per bit compared to switching; (5) Energy estimates depend on the metrics used and hence developing appropriate benchmarks are critical to understanding and quantifying the energy partitioning between the different computing processes in systems; (6) Nature still appears to be more efficient in information processing and is operating closer to thermodynamical limits than computing systems [30].

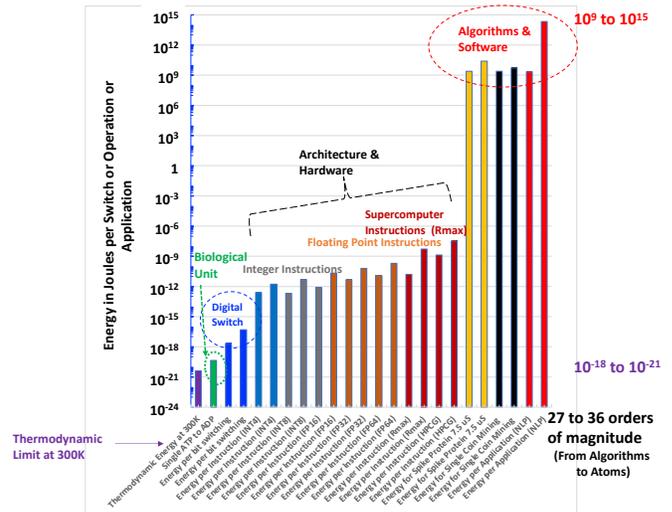

**Figure 5**: Energy for different layers of computing with lower and upper bounds as estimates compared with the physical (thermodynamics-based) and biological (metabolism-based) limits. Moving from atoms and molecules (thermodynamic and biological entities) to training machine learning for large language models, high performance scientific simulations, and Crypto coin mining, 27 to 36 orders magnitude are observed.

As had been indicated before [3], energy efficiency due to geometrical scaling is slowing down due to increasing complexity in design, processing, integration, and manufacturing. Given this slowing, energy efficiency needs to be reached through multiple innovations spanning Application-level computations with bit-level switching. Since the energy efficiency at the bit-level does not proportionately translate to the instructions at the system level, it is important that systematic design strategies connecting atoms (switches) with architecture/hardware and algorithms/software need to be adopted for a new era in sustainable computing.

Our analysis of three different simulation examples indicate that Application-level efficiency needs to be an additional factor in system-level design. With increasingly powerful computers reaching energy limits driven by power constraints and refrigeration requirements, it is important that efficiency in the algorithm/software need continued focus as well. Based on the analysis presented, it can be concluded that the large orders of magnitude in energy estimates indicate that computing may need to be reframed with lessons to be learned from Nature on efficient processing of information [31-32].

ACKNOWLEDGEMENTS

This work was partially supported by the U.S. Department of Energy's Office of Science contract DE-AC02-76SF00515 with SLAC through an Annual Operating Plan agreement WBS 2.1.0.86 from the Office of Energy Efficiency and Renewable Energy's Advanced Manufacturing and Materials Technology Office. The institutional support from SLAC National Laboratory, the US DOE's EES2 Working Groups, and MIT LLSC for their inputs (A. Reuther, J. Kepner) are also acknowledged.